\documentclass{article}
\usepackage{fullpage}
\usepackage{amsmath,amssymb,amsfonts}
\usepackage{graphicx}
\usepackage{cleveref}
\usepackage{url}
\usepackage{xcolor}

\usepackage{cleveref}

\begin{document}

\section*{Quantitative errors in the Cochrane review on ``Physical interventions to interrupt or reduce the spread of respiratory viruses"} % Article title

\begin{center}
Yaneer Bar-Yam\textsuperscript{1}, Jonathan M. Samet\textsuperscript{2},  Alexander F. Siegenfeld\textsuperscript{1,3}, Nassim N. Taleb\textsuperscript{4} \\
\textsuperscript{1}\textit{New England Complex Systems Institute}, \textsuperscript{2}\textit{Colorado School of Public Health}\\
\textsuperscript{3}\textit{MIT}, \textsuperscript{4}\textit{Tandon School of Engineering, New York University, and Universa Investments} \\
\end{center}

{\bf The COVID-19 pandemic has brought a heightened sense of urgency in the scientific community regarding the need to advance understanding and prevention of pathogen transmission, particularly concerning infectious airborne particles and the utility of various preventive strategies in reducing the risk of infection. There are extensive studies validating scientific understanding about the behavior of larger (droplets) and smaller (aerosols) particles in disease transmission and the dosimetry of particles in the respiratory track. Similarly, modalities for respiratory protection against particles in the size range spanned by infectious particles, such as N95 respirators, are available and known to be efficacious with tested standards for harm reduction across environments including physical, chemical and biological hazards. Even though multiple studies also confirm their protective effect when adopted in healthcare and public settings for infection prevention, overall, studies of protocols of their adoption over the last several decades in both clinical trials and observational studies have not provided as clear an understanding. Here we demonstrate that these studies are strongly biased towards the null by infections resulting from transmission outside of the investigated environments and study participants. Such study limitations are frequently mis-stated as not influencing the conclusions of research on respiratory protection. The reason for the failure to properly analyze the studies is that the standard analytical equations used do not correctly represent the random variables that play a role in the study results. By correcting the mathematical representation and the equations that result from them, we demonstrate that conclusions drawn from these studies are strongly biased and much more uncertain than is acknowledged, providing almost no useful information. Even with all these limitations, we show that existing results, when outcome measures are properly analyzed, consistently point to the benefit of precautionary measures such as N95 respirators over medical masks, and masking over its absence. We also show that correcting manifest errors of widely reported meta-analyses also leads to statistically significant estimates. Our results have implications for the design of studies and analyses on the effectiveness of respiratory protection and on using existing evidence for policy guidelines for infection control.}

\section{Introduction}

The COVID-19 pandemic brought urgency to advancing evidence on the clinical and public utilization of preventive interventions for airborne pathogens, with a specific focus on mask wearing, particularly N95 respirators, which protect their wearers against inhaling small and large particles and also reduce particle emissions from their wearers. While the effectiveness of respiratory protection through the filtration of airborne particles in inhaled air is well-established, there have been a limited number of randomized clinical trials on their use in real-world settings. These trials and observational studies have not consistently shown beneficial outcomes in evaluations of specific protocols for using respiratory protection. As a result, controversies have emerged regarding the appropriate interpretation of these findings. For example, a recent Cochrane systematic review and meta-analysis examining the effectiveness of ``Physical interventions to interrupt or reduce the spread of respiratory viruses" \cite{R1} has faced significant criticism \cite{R2,R3,R4,R5} for its interpretation of the evidence with the suggestion that masks are ineffective and that N95 respirators and surgical masks are equivalent in protection afforded. It is important to note that the studies reviewed have recognized limitations and inherent uncertainty. The editor-in-chief of Cochrane voiced concerns regarding the presentation and interpretation of this review \cite{R6}. A systematic review without a meta-analysis covered studies on prevention of SARS-CoV-2 only.  That review identified only a few studies and offered a similar set of findings, while acknowledging methodological limitations of the studies considered (Chou and Dana 2023 \cite{ChouDana2023}). Long 2020 \cite{R8} reported statistical significance for the benefit of N95 respirators versus medical masks in sensitivity analyses of meta-analyses when removing just one of their included studies in two different observed endpoints, Loeb et al 2009 \cite{R9} for prevention of laboratory-confirmed respiratory viral infections, and Radonovich et al 2019 \cite{R10} for prevention of respiratory infections. These reviews demonstrate the limitations of the studies for offering conclusive findings but the methodological issues that lead to uncertainty have not been clarified.

Here, we explore the methodological challenges in investigations of the efficacy and effectiveness of studies of respiratory protection and risk for infection. We find that these challenges have not been adequately taken into account in the design of specific trials and observational studies and that their consequences spill over to systematic reviews and meta-analyses.  In particular, there are strong biases towards the null affecting both superiority studies, i.e., are N95 respirators superior to surgical masks and other barrier face coverings, and to equivalency studies, i.e., are N95 respirators and surgical masks and other barrier face coverings equivalent in their effects. Given the need for evidence-based policy determinations to protect workers and the public, the findings of such studies, and of systematic reviews and meta-analyses, in particular, have substantial policy implications.   

We describe these methodological challenges in estimating the effect of respiratory protective measures against transmission of airborne pathogens.  Specifically, we provide the underlying mathematics for exposure misclassification and its consequences in individual studies and show the implications of the propagation of the associated uncertainty in systematic reviews and meta-analyses. We focus on the Cochrane review and the recent trial by Loeb et al \cite{R11}. Our findings have broad implications for study design, and interpretation of the collective body of trial and observational evidence on the beneficial effects of N95s and other personal respiratory protective devices. 

Our identification and analysis of methodological challenges in the studies of respiratory protection exemplifies a widespread challenge in statistical science whose discovery and resolution in statistical physics caused a revolution in the study of phase transitions \cite{rg} that expanded to many other applications \cite{kardar} and multiscale analysis \cite{bigdata,multiscale} (for an introduction see \cite{whycomplexity}). An explanation of the essential concepts of this framework involves relevant variables and universality which specifies the mathematical representation to be used for analysis of a system. Limitations to phenomenological approaches \cite{limits} include the need for adequate characterization of the system under study. Our analysis here reflects the general conclusion that the relevant variables must be identified prior to either valid or validatable analyses.

\section{Methodology of methodological analysis}

  \begin{figure}
    \centering
    \includegraphics[width=.80\textwidth]{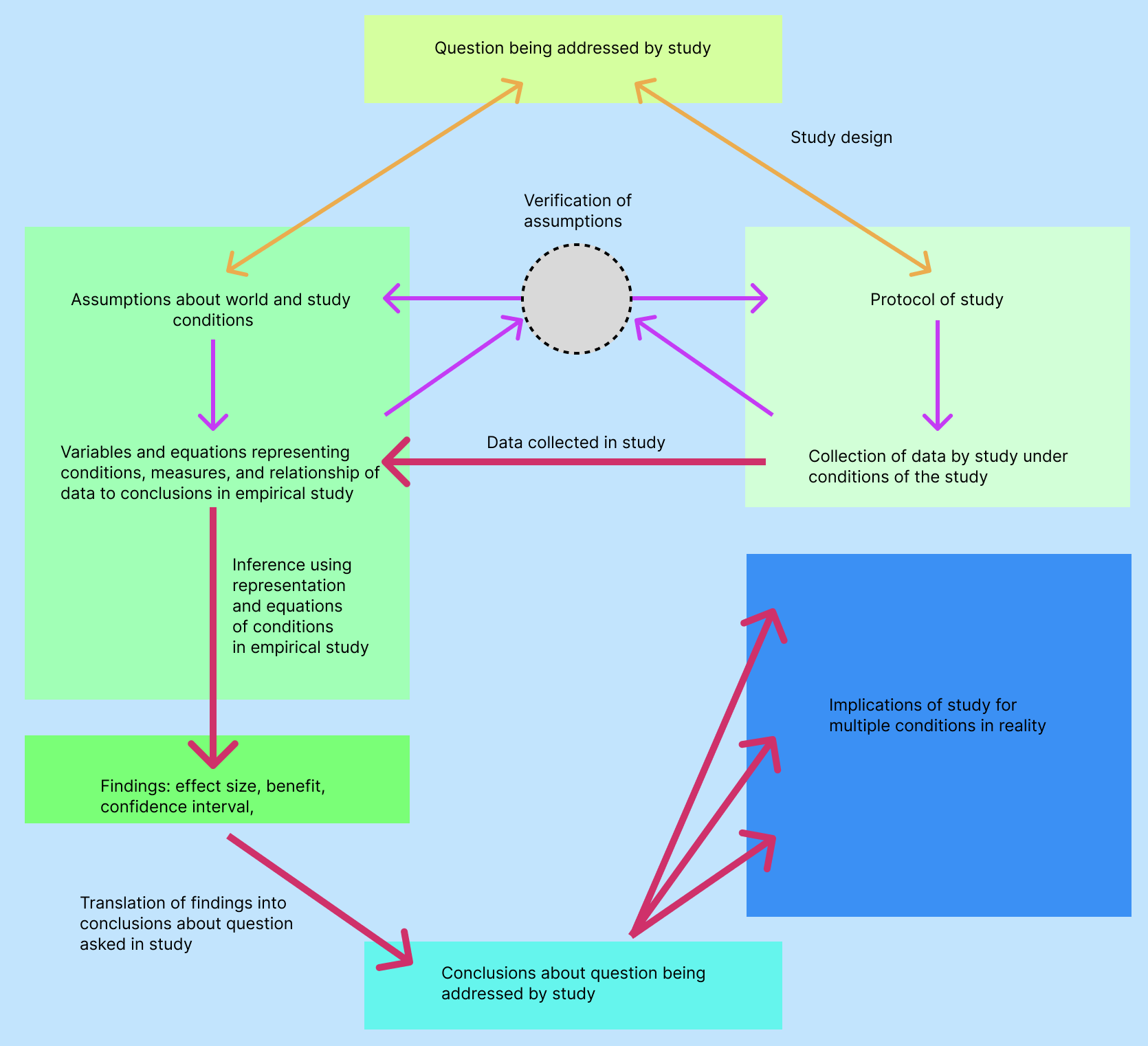}
    \caption{In an empirical study, there is no path directly from the data to the conclusions and its implications for conditions in the real world. The evidence provided by data is derived through a set of equations that provide effect estimates and confidence intervals to infer findings. The results that these equations provide (estimates of effect size, confidence intervals, and measures of fit) are translated into conclusions about the question asked by the study. The representation used (variables) and the equations always reflect assumptions that must be verified for a particular study condition. This framing was not done adequately for studies and meta-analyses discussed in this paper leading to large differences between the conclusions inferred  by the authors and others and those that would be inferred based upon corrected equations (see Fig. 2).}
    \label{fig:theoryexp}
\end{figure}

  \begin{figure}
    \centering
    \includegraphics[width=.99\textwidth]{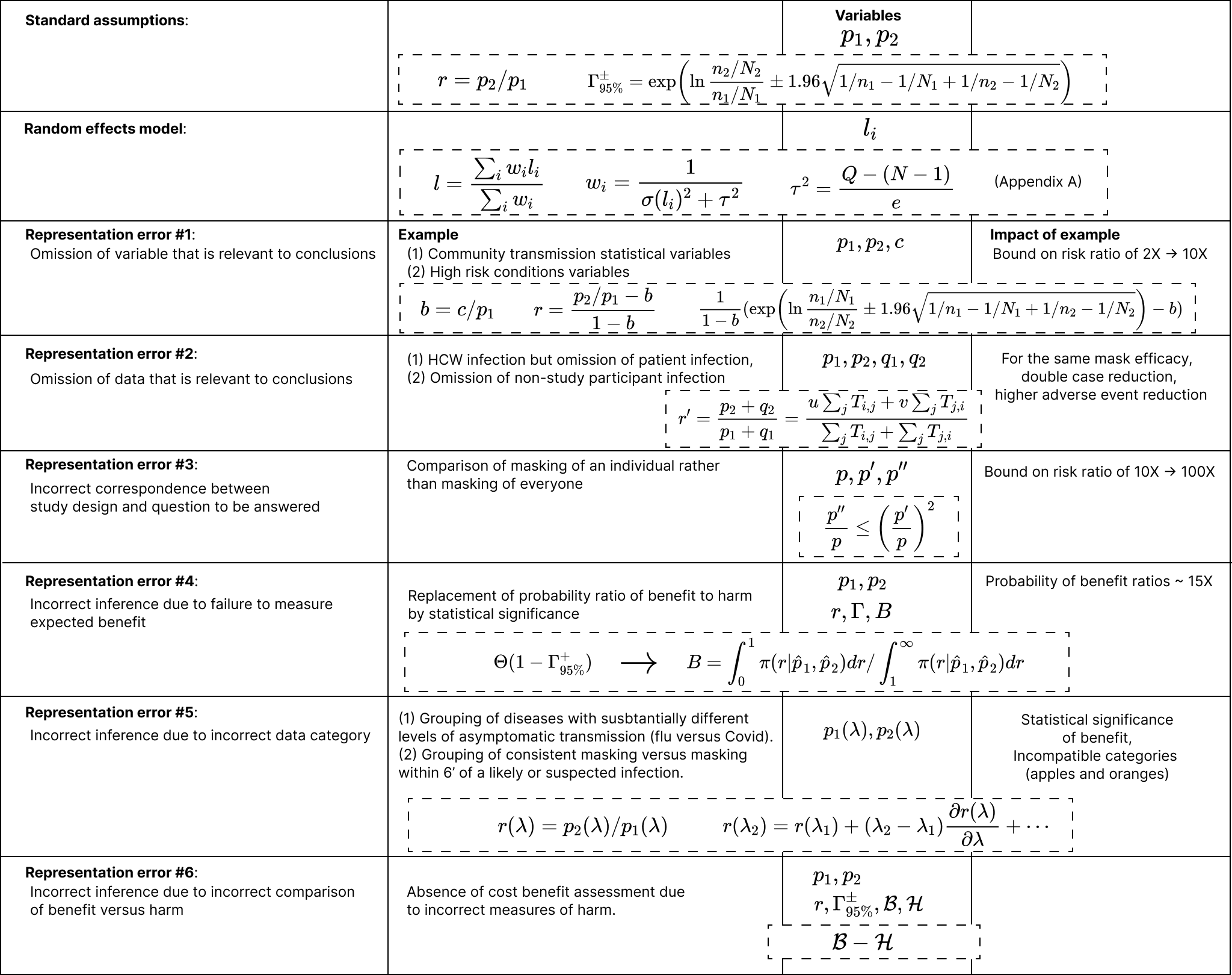}
    \caption{A failure to evaluate assumptions and incorporate mathematical equations that correspond to study conditions leads to incorrect conclusions. The first row shows a standard representation of estimates of effects and confidence intervals. In many randomized controlled trials (RCTs) and meta-analyses, random effects equations (second row and Appendix A) are used to account for the impact of unobserved effects that result in random differences between study groups, but do not address systematic biases. Rows 3 through 8 in the table show six distinct types of errors in RCTs and meta-analyses of mask efficacy. Examples of how they appear in such studies are described in the text, including changes in variables, corrected equations for results (dashed rectangles) and consequences for conclusions. Conclusions revised based on these corrections are dramatically different from those  drawn from individual studies and meta-analyses. The differences extend to estimates of effect size, confidence intervals, statistical significance, and the magnitude of benefit (see text).}
    \label{fig:math}
\end{figure}

In this paper we reconsider the validity and necessary corrections of the methodology of multiple studies, including randomized trials and associated meta-analyses that have been used to evaluate the prevention of pathogen transmission through respiratory protective devices, including N95 respirators and medical masks. Our meta-methodology---methodology for this methodological analysis---is to analyze the correspondence between the study conditions, the mathematical analysis of the study data, and the question being addressed by the study. The conditions of a study must be reflected in the mathematical representation and equations used to analyze the data from the study for valid outcome measures to be obtained and for the analytical results to be correctly interpreted (see \cref{fig:theoryexp}). For example, the standard equations for risk ratios and confidence intervals in the top line of \cref{fig:math} are obtained from a two variable independent event model comparing two conditions by the frequency of events in two populations under protocols that embody those conditions. However, these assumptions do not generally hold for empirical studies. An example of corrections to this representation is the common assumption of a random effects model for meta-analyses and individual studies that include multiple sub-populations whose conditions vary in unspecified ways and are assumed in their analysis to have normally distributed random effects (see line two of \cref{fig:math} and Appendix A, including the use of log normal distributions for risk ratios). These equations become necessary when standard equations are not valid for a study condition. This is not, however, the only change that may be needed. 

We show there are multiple corrections needed for correspondence between the study methods and the data analysis---because neither the standard nor the random effects mathematical representation correspond to the conditions of the studies performed. Instead, systematic biases occur (i.e., non-random effects) that modify the mean effect sizes and the confidence interval and thus also the inferred study outcomes and conclusions.  Our methodology identifies missing variables and data, and other inconsistencies between study conditions and the analyses of study data that are carried out to address study questions and to reach conclusions, as summarized in \cref{fig:math}. We show that these biasing effects are large and neglected in analyses of the data. Quite generally, studies must use equations derived from validated assumptions to qualify as valid and useful empirical studies. This can be achieved by various combinations of modifying the study conditions, modifying the mathematical equations, or modifying the question being asked. 

We note that the missing variables we identified do not fall under the category of ``confounding variables," which are variables that are associated with the independent variables of interest and with the outcome variable.  Such variables distort the association between variables of interest and the outcomes. Instead, our analysis highlights the variables that are essential for accurately characterizing the study conditions, also known as relevant variables.

We show that the analyses of randomized clinical trials are missing six things: (1) propagation of uncertainty from improperly neglected random variables, (2) compounding of effects due to unaccounted transmission  and infection of non-study participants, (3) invalid correspondence between study question and design reflected in variables used, (4) analysis of significance---the meaning of the results through their implications for health, (5) invalid categorization of data associated with study conditions, and (6) clear definitions and characterization of adverse effects. Using two recent reports---a trial and a systematic review and meta-analysis, we show that studies comparing N95 respirators and surgical masks, while interpreted as showing equivalency, are compatible with a substantial benefit of N95s. 

It is crucial to acknowledge the significant impact that correcting the methodology of these studies can have. For the trial by Loeb et al. we find that the study data are consistent with an upper bound of benefit for N95 respirators compared with surgical masks of 10X or larger rather than less than 2X.\footnote{Confidence intervals are much wider than reported, even unbounded.} We further show that the results only apply to an unspecified class of low risk procedures; that including the effect of missing data on source control on infections of others (including patients and non-study participant healthcare workers) results in a much higher limit on harm caused; that two-way masking leads to an upper bound of benefit of 100X or larger; and that reports of over $10\%$ ``adverse events" are non-severe (mild) ``discomfort, skin irritation, and headaches." Finally, we show that existing data indicate with high probability---a 15X odds ratio---that effective masking leads to improved health outcomes. Our methodology (detailed in Appendices) and results confirm previous analysis that when the assumptions of studies are corrected, the available empirical evidence is consistent with masks being effective \cite{R7}. 

The corrections of existing studies' conclusions using our meta-methodology show that there is no such thing as a purely empirical result; all empirical studies require theory with which to interpret them. Thus, even when conducting empirical research, careful attention must be placed on the implicit theoretical assumptions used to interpret the data. Ignoring the validity of the assumptions used in the analysis of empirical data is misleading and, in the case of the Cochrane review, leads to incorrect results. A key example is the neglect of ``propagation of uncertainty'' in analyzing study data (for an introduction see \cite{propagationofuncertainty}). 

\section{Background}

In a standard model of a study of interventions to reduce incidence risk, individuals are divided into two groups and subjected to two conditions $A_1$ and $A_2$. The ratio of incidence is used as the comparison metric. By comparing the incidence rates, denoted as $p_1$ and $p_2$, in the two groups containing $N_1$ and $N_2$ individuals, respectively, conclusions are drawn. These conclusions are derived based on standard statistical assumptions that consider the association between conditions $A_1$ and $A_2$ and the occurrence of the outcome measures $p_1$ and $p_2$ during conditions $A_1$ and $A_2$. The true values of $p_1$ and $p_2$ are unknown but are considered representative of the conditions of each group. In a typical statistical analysis, efforts are made to estimate the risk ratio using the equation:
\begin{equation}
r = p_2/p_1  
\label{eq:r}
\end{equation}
Additionally, standard approximate formulas are used to determine the upper and lower bounds of a 95\% confidence interval, which can be represented as:
\begin{equation}
\exp\left(\ln\frac{n_2/N_2}{n_1/N_1}\pm 1.96\sqrt{1/n_1-1/N_1+1/n_2-1/N_2}\right)
\label{eq:x}
\end{equation}

When applying this model to a real-world study, a challenge arises in that the intended exposure conditions $A_1$ and $A_2$ may not align with the actual circumstances of the study. For example, if a study aims to draw comparative conclusions about conditions $A_1$ and $A_2$, it may implement comparison and intervention protocols $B_1$ and $B_2$ in two groups. While conclusions using the above expressions might be made about protocols $B_1$ and $B_2$ as they are implemented, these conclusions may not correspond to conditions $A_1$ and $A_2$. To address this issue, various corrections may be applied to infer conclusions about conditions $A_1$ and $A_2$ if they are adequately analyzed and tested.

We address the imperfect association between protocols $B_1$ and $B_2$ and conditions $A_1$ and $A_2$ by considering a larger set of conditions reflective of the actual conditions of such studies. Specifically, we examine a scenario where the target conditions $A_1$ and $A_2$ are realized at certain times for study participants; however, the protocols do not ensure that target conditions are consistently followed across individuals over time resulting in contributions to incidence rates that are not associated with the target conditions. This context is illustrated in Table 1, which presents a two by two incidence matrix for target conditions $A_1$ and $A_2$, and non-target conditions, i.e., outside of the study’s spatio-temporal domain, that can contribute to incidence observed during both of the protocols.

\begin{table}[h!]
\begin{center}
  \begin{tabular}{@{} |c|c|c| @{}}
    \hline
    Study Group & Target Incidence & Non-Target Incidence \\ 
    \hline
    Comparison & $I_{1t}$ & $I_{1c}$ \\
    Intervention & $I_{2t}$ & $I_{2c}$ \\
    \hline
  \end{tabular}
\end{center}
\caption{The two by two table provides the incidence rates of interest in a study of the efficacy (trial) or effectiveness (observational study) of an intervention to reduce risk of infection from an airborne pathogen. The table is stratified by the incidence rates for infection by study group, intervention or comparison, and by the place where infection occurs, i.e., the targeted environment such as a healthcare facility, and in the other places where infection might occur, subscripted by $c$, referring generally to “other” for the non-targeted environments.  To test the effect of the intervention, the desired estimand is the relative risk, i.e., the ratio $I_{2t}/I_{1t}$.  That ratio cannot be directly estimated since the location where infection is acquired is not knowable.  In actuality, the relative risk is estimated as the ratio $(I_{2t} + I_{2c})/(I_{1t} + I_{1c})$.  While $I_{1c}$ and $I_{2c}$ may be comparable, the estimated ratio deviates from the desired estimate of relative risk with the magnitude of the deviation depending on the incidence rates in the places where the intervention is not employed.  }
\label{table:1}
\end{table}

\section[A microenvironmental model for infection describing omitted variables relevant to study conclusions]{A microenvironmental model for infection describing omitted variables relevant to study conclusions} 

\subsection{A microenvironmental model for infection}

For mask studies, the conditions studied are either the wearing and not wearing of a mask, or the wearing of two different kinds of masks, often a medical/surgical mask versus an N95 or other respirator mask. The study is typically performed by providing the protective device and guidelines for wearing it, but whether the guidelines ensure the desired masking behavior must be evaluated.

We introduce a straight-forward micro-environmental exposure model for risk of respiratory infection (Figure 1).  In this model, people spend time each day in various microenvironments, each having a particular risk for infection during the time spent there.  Risk for an individual reflects the risk cumulated across the various microenvironments with their associated exposures to pathogens and corresponding risks for infection.  For many reasons, risks would be expected to vary across microenvironments. Figure 1 extends the microenvironmental model to the circumstances of a trial.  For trials of respiratory protection involving health care or other workers, the intervention is typically applied only to the work environment while the outcome is incidence of infection as it reflects exposures in all microenvironments.  The microenvironment where exposure leads to an infection in a study participant is not knowable.  In fact, depending on the array of microenvironments, the study environment, e.g., a healthcare setting, may not be the dominant contributor to risk for infection.

The standard analysis has been stated to rest upon the assumption that the infection risks outside of the target microenvironments are comparable in the intervention and comparison groups.  While perhaps correct in practice, albeit unverifiable, analytical approaches to deriving conclusions from the study through evaluation of the risk ratio and confidence interval do not incorporate the uncertainty that arises from infections resulting from exposures in the microenvironments not targeted by the intervention.  In fact, depending on the circumstances, the risk for infection associated with microenvironments other than the intervention microenvironment could be the dominant contributor to risk for participants in a study.  

Unmasked exposures in the intervention group are often mentioned as a limitation to these studies, but no attempt is made to modify the mathematical analysis accordingly; rather, a \textit{pro forma} structure for the research protocol and its analysis appears to be followed, which gives a superficial appearance of scientific validity without conforming to scientific standards by appropriately reflecting bias and uncertainty.

\subsection{Exposure misclassification and its consequences}

\begin{figure}
    \centering
    \includegraphics[width=.60\textwidth]{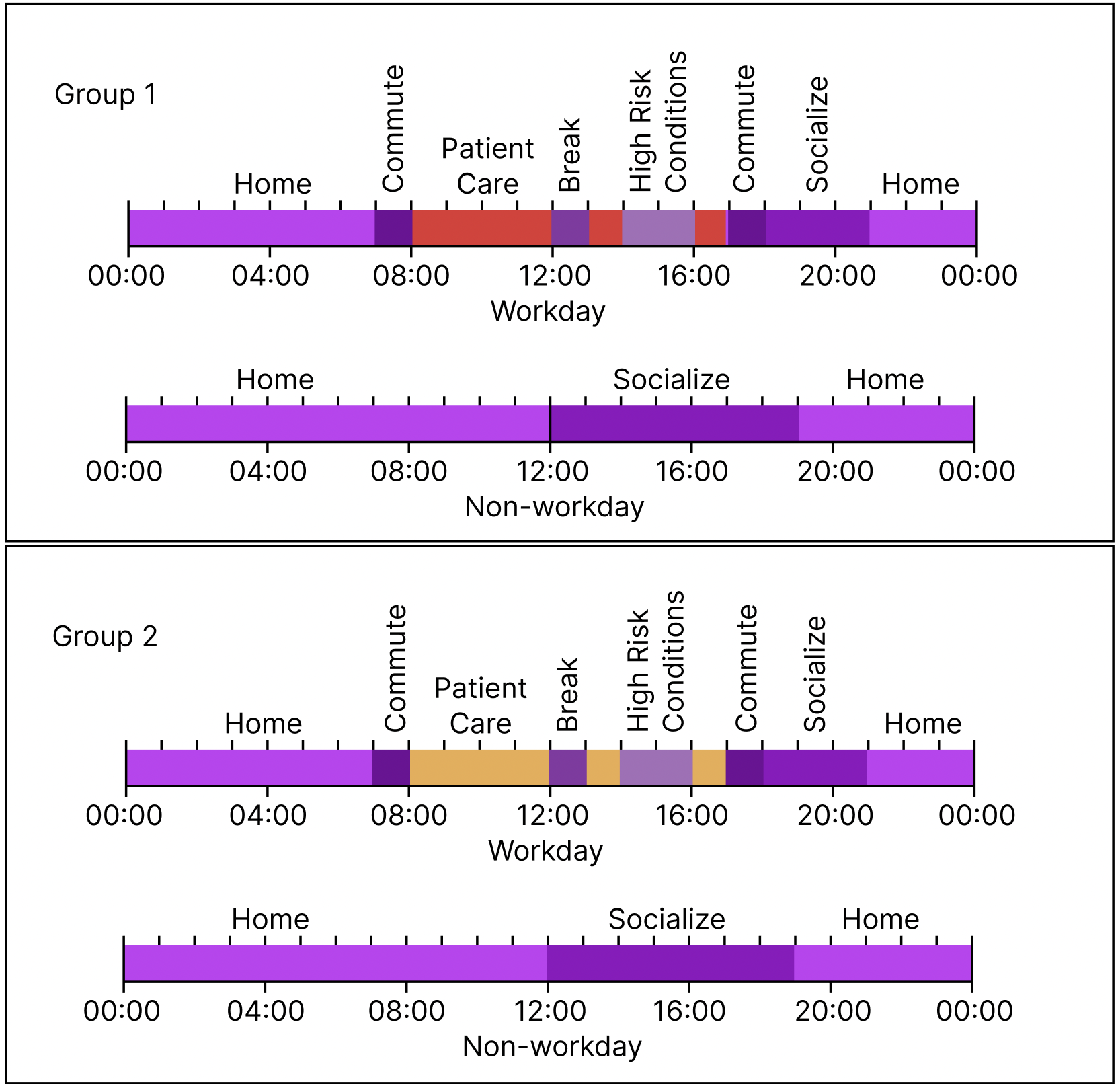}
    \caption{Schematic diagram of the daily activities of members of two groups according to the protocols of studies investigating healthcare provider masking behavior. Each individual encounters, or by behavioral choices engages, in multiple micro-environments. Protocols control only masking behavior during some of them, particularly patient care. Some studies also prescribe specific common masking behavior for both groups, e.g. nonspecific ``high risk conditions.'' The times that are distinguishing the two groups are the yellow and red periods, all the shades of purple including home, commute, socializing, and, where specified, high risk conditions, are common between the two groups. These times all contribute and aggregate into usually unaccounted for noise when the analysis of the study is done. See text for an analysis that clarifies their contribution which leads to bias in point estimates and uncertainty intervals.}
    \label{fig:microenv}
\end{figure}

\noindent Here, we consider the consequences of what we term “exposure misclassification”, i.e., the time at risk for infection that is not covered by the intervention.  There are implications for both the estimate of effects and the statistical uncertainty of the estimate.  In particular, as described above, participants in studies pass through micro-environments that differ over time for study participants as illustrated in \cref{fig:microenv}. Often, per study protocols, many micro-environments do not have proscribed protection with the study intervention. For example, protocols may be designed to only control the difference between groups during a particular period of time, i.e., healthcare workers treating patients who are confirmed or suspected to be infected with a respiratory virus, and excluding periods where they are engaged in high risk activities. However, time spent at home, commuting, socializing, and, in specified, high risk conditions is not controlled, Additionally, study protocols may be limited in the extent to which they influence participants’ compliance with the protocols. If controls are not present during these periods, the results of a study must be analyzed accordingly. For instance, even if masks provide perfect protection, $p_2/p_1$ will still be limited by study design if consistent mask use during the period of the study was not ensured. Moreover, the confidence interval given by \cref{eq:x} is for $p_2/p_1$, which is a distinct quantity from the protection that masks actually provide under the conditions of the study.
 
  \begin{figure}
    \centering
    \includegraphics[width=.45\textwidth]{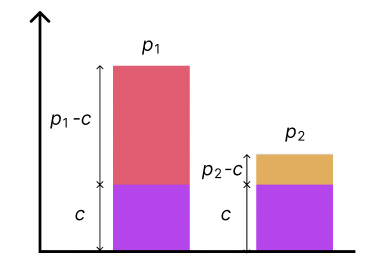}
    \caption{The expected fraction of infections $p_1$ and $p_2$ in each arm of a study are shown. If in each arm a fraction $c$ were infected due to exposures outside of the study microenvironment, then the true risk ratio is given by $r = (p_2 - c)/(p_1 - c)$. See Appendix B for more details.}
    \label{fig:diagram1}
\end{figure}

\subsection{Modification of confidence intervals to incorporate exposures outside of the study}
We can modify the point estimate and confidence intervals obtained above in order to account for these unprotected exposures in the intervention and comparison groups.  Continuing with the notation above and letting the additional random variable $c=n_u/N_2$ where $n_u$ is the expected value of the number of infections arising from exposures in the intervention and comparison groups, outside of the intervention microenvironment:
\begin{equation}
p_2-c=r(p_1-c)
\label{eq:c}
\end{equation}
where $r$ is the actual risk ratio. In other words, as shown in \cref{fig:diagram1}, the expected fraction of infections from exposures in the intervention microenvironment ($p_2-c$) will differ from the expected fraction of infections from the corresponding exposures in the control group ($p_1-c$) by a factor of $r$ (see Appendix B for a more detailed derivation that accounts for heterogeneity and multiple exposures).

Then, defining $b=c/p_1$ yields an expression for the risk ratio:
\begin{equation}
r=\frac{p_2/p_1-b}{1-b}
\label{eq:sb}
\end{equation}
The confidence intervals given by the meta-analyses are for the value of $p_2/p_1$, and so we can use \cref{eq:sb} to calculate the confidence interval for the actual risk ratio, subject, of course, to both the assumptions here and the assumptions of the meta-analysis, except for this modification.  For instance, applying \cref{eq:sb} to the approximate formula given in \cref{eq:x} yields the following confidence interval estimates:
\begin{equation}
\frac{1}{1-b}(\exp\left(\ln\frac{n_2/N_2}{n_1/N_1}\pm 1.96\sqrt{1/n_1-1/N_1+1/n_2-1/N_2}\right)-b)
\label{eq:cib}
\end{equation}
But more generally, \cref{eq:sb} can be applied to any estimate of $p_2/p_1$, including both the point estimate as well as the upper and lower bounds of the confidence interval, regardless of how they were calculated, in order to obtain a corresponding estimate for the actual risk ratio $r$.

\begin{figure}
    \centering
    \includegraphics[width=.48\textwidth]{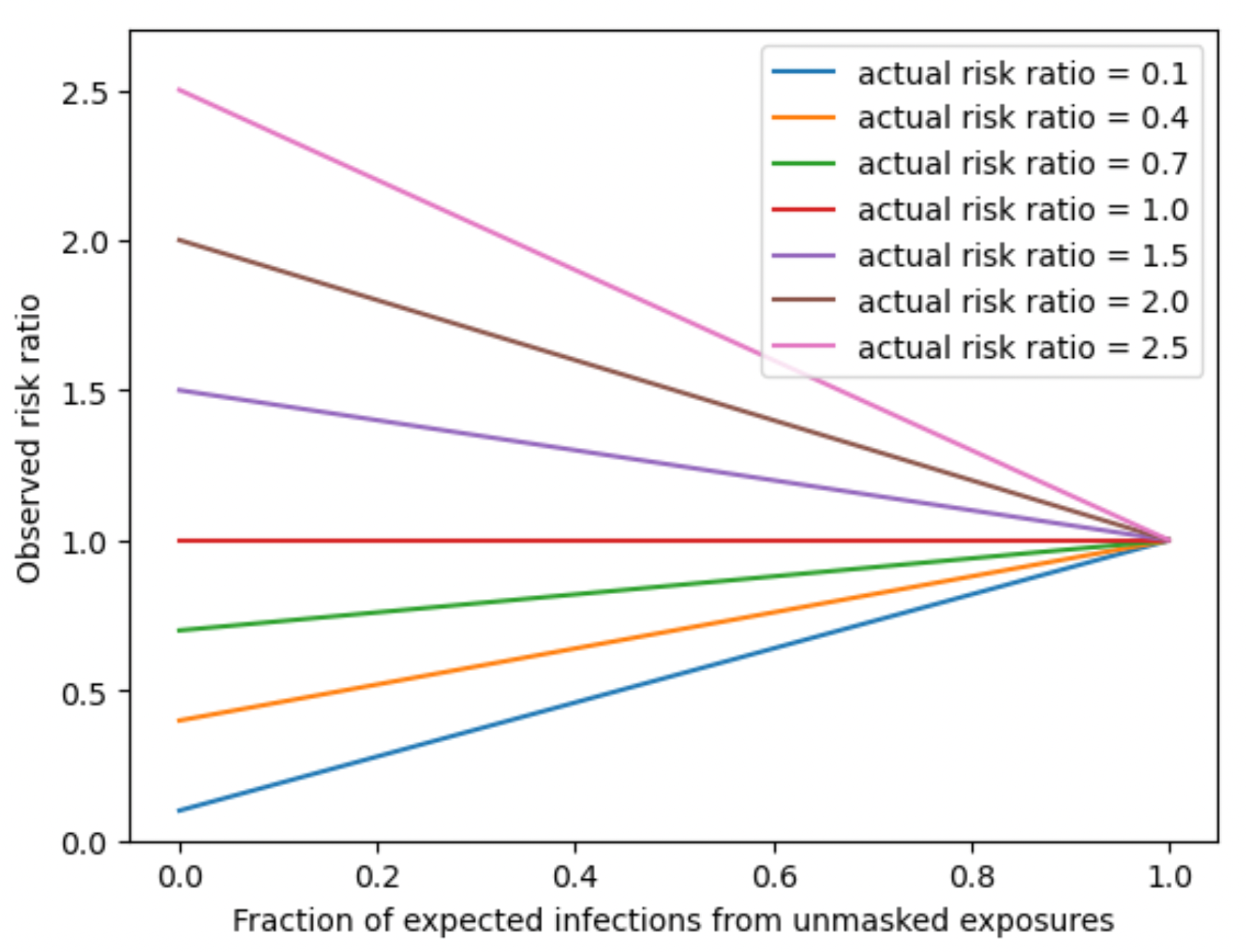}
    \includegraphics[width=.48\textwidth]{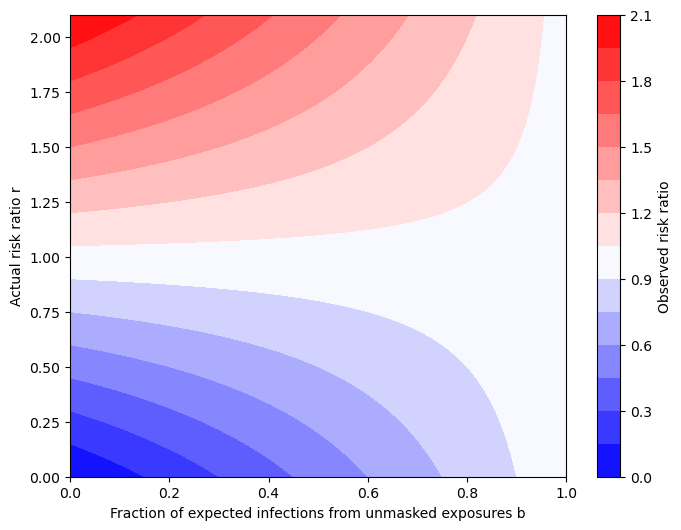}
    \caption{
    The value of the risk ratio generating the observed data ($p_2/p_1$) is shown for various potential values of the actual risk ratio $r$ and the fraction of unmasked exposures $b$. Note that as the fraction of unmasked exposures increases, the observed risk ratio will approach 1, regardless of actual efficacy. Thus, if the fraction of risk from unprotected exposure is relatively high, a small uncertainty in the observed risk ratio corresponds to a much larger uncertainty in the actual risk ratio. For small or large risk ratios (corresponding to masks having a large effect on transmission), changing the fraction of unmasked infections has a substantial effect, as compared with risk ratios close to $1$ where there is very little difference. This sensitivity analysis shows that such studies underestimate the effectiveness of high quality masks, systematically shifting the results toward a conclusion of ineffective masking. }
    \label{fig:theory}
\end{figure}

\begin{figure}
    \centering
    \includegraphics[width=.45\textwidth]{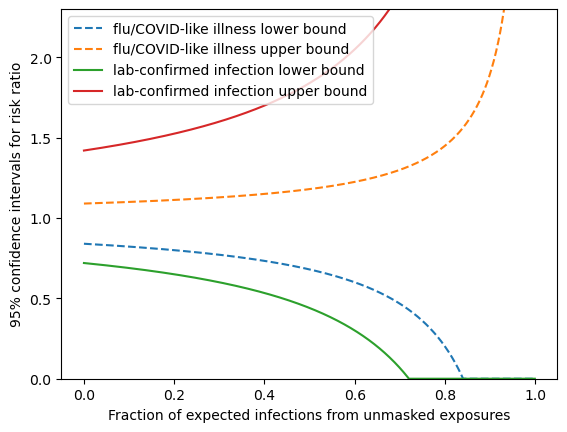}
    \includegraphics[width=.45\textwidth]{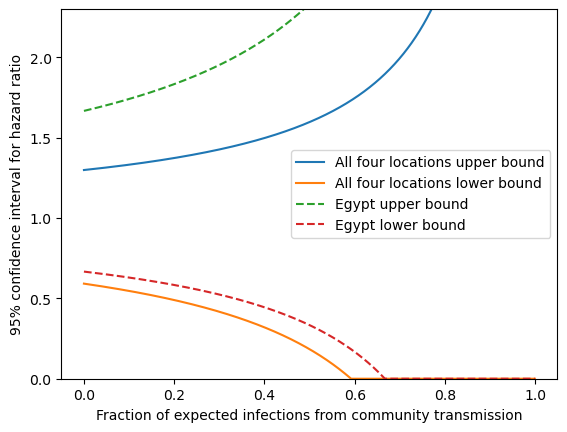}
    \caption{Left panel: Upper and lower bounds for the confidence intervals of the risk ratio (a risk ratio of zero would mean masks are perfectly effective, while a risk ratio greater than one would mean they are harmful) of surgical masks vs. no masks as a function of $b$ (\cref{eq:sb}), the expected number of infections from exposures while participants were unmasked as a fraction of the total expected infections if no masks are worn. Right panel: Upper and lower bounds for the confidence intervals of the hazard ratio of N95 respirators vs. surgical masks (i.e. the ratio of the probability of infection while wearing an N95 respirator to the probability of infection while wearing a surgical mask at any given time) as a function of $b$ (\cref{eq:nb}), the expected amount of infections from outside the trial setting as a fraction of the expected total number of surgical mask group infections.}
    \label{fig:surgicalandN95}
\end{figure}

We see in \cref{fig:theory} that as unmasked exposures increase, an increasingly large range of actual risk ratios will result in an observed risk ratio close to $1$, producing a systematic bias towards the null and smaller estimates of effect.  Because of this bias, a small degree of uncertainty in the observed risk ratio corresponds to a much wider range of uncertainty in the actual risk ratio. Thus, accounting for these unmasked exposures results in a widening of previously reported confidence intervals. In the left panel of \cref{fig:surgicalandN95} we show the upper and lower bound of the $95\%$ confidence intervals for two meta-analyses considering surgical masks vs. no masks (influenza/COVID-like illness and laboratory-confirmed influenza or SARS-cov-2) as a function of b in the Cochrane review.

We note that $b$ will be larger than the percentage of exposures for which people are unmasked (see Appendix B).  
For instance, if masks are worn for 30\% of exposures (and for many of the studies, this is likely an overestimate due to the protocols used), $b$ could easily be larger than $0.7$ (i.e. $1-30\%$) because many of the individuals exposed while masked would have been infected anyway from their unmasked exposures.  
Given that values of $b$ close to $1$ cannot be ruled out, the left panel of \cref{fig:surgicalandN95} indicates that neither very high nor low levels of mask effectiveness are inconsistent with the data analyzed by the Cochrane review; in other words, the data included in the meta-analysis are uninformative one way or another. 

\subsection{Consequences: Comparison of N95 respirators with surgical masks}

To further illustrate the quantitative consequences of misclassification, we consider a study on N95 vs. surgical masks \cite{R11} (Loeb 2022) that was mentioned but not included in the Cochrane review.   A number of published comments have criticized the study by Loeb 2022  \cite{R12}; here, we use a similar analysis as above to demonstrate the compromise of its conclusions by the methodological problem explored above.  The study conclusions claim a limit of risk ratio $p$ of $2$X higher from surgical masks, i.e., the noninferiority risk ratio. Our analysis indicates it is unbounded by the study. 

The study reports on results obtained by applying the same protocol in four locations; the results are dominated by infections from the fourth location (Egypt), for which results were obtained during the Dec 2021-Feb 2022 Omicron wave with widespread public infection.  The data are interpreted to show that there is no difference between those who were protected with N95 and those using a surgical masks in the hospital setting.  But the hospitals were not the only high risk setting for those workers during that period of time. The high rates of transmission during data collection in Egypt may have biased this component of the study more strongly towards the null than the results from the other three locations.

When community transmission is considered (Appendix C), we obtain the bounds for the hazard ratio\footnote{For the sake of consistency with our analysis of surgical masks vs. no masks, the hazard ratio given here is for N95 respirators to surgical masks, which is the inverse of the hazard ratio for surgical masks to N95 respirators given in the original study.} shown in
right panel of \cref{fig:surgicalandN95}. While the study acknowledges community transmission as a limitation and performs a post-hoc subgroup analysis\footnote{The subgroup analysis (shown in Supplement Table 3 of ref.~\cite{R11}) finds no statistically significant difference between the hazard ratios of those who reported known community/household exposure and those who did not.}, it fails to modify its confidence intervals to account for community transmission, despite the  data indicating that over $50\%$ of the participants reported known exposure to community/household respiratory illness (see Supplement Table 3 of ref.~\cite{R11}). Additionally, many community/household exposure events may not have been known (and/or not reported), while the reported infection rate was under $15\%$. The near equal infection rates are consistent with a similar and much higher exposure rate in the community providing an undetermined but possibly high value of $b$. Counter to the authors’ conclusions, correction for infection outside of the intervention microenvironment shows that the study provides no information relevant to the study question, i.e., the range of compatible estimates is expansive (the limit is infinite, i.e. the inverse of what is plotted in the right panel of \cref{fig:surgicalandN95}, which can be zero) on the relative risk of surgical masks versus N95 respirators. Thus, the actual bound on risk from surgical masks from this study is not less than 10X that of N95 respirators and could be higher. 
 
The study protocols also specified that ``participants who believed they were at high risk during a particular exposure were allowed to use the N95 respirator if assigned to a medical mask".  This provision, while warranted for worker protection, has implications as any infections resulting from these unspecified high risk conditions will increase the parameter $b$ (the same parameter that accounts for infections from community transmission). If the actual risk ratio in high risk conditions is the same as the risk ratio in low risk conditions, then the subtraction of those cases through an increase in $b$ will give the same risk ratio as if the infections were not included because of the N95 masking. Just as in the case of community transmission, these increases in the parameter $b$ due to high risk conditions will systematically bias the study's results towards smaller effects (\cref{fig:theory}), and will widen the true confidence intervals (\cref{fig:surgicalandN95}). This potential limitation was not addressed by the authors. 

Other studies in the Cochrane review that compare surgical masks with N95 respirators suffer from similar flaws. While some of these studies mention community transmission as a limitation, we find an underlying misconception that if community transmission is approximately equal in both arms of a study then there is no basis for concern about bias.\footnote{Loeb 2009 \cite{R9} states ``It is also impossible to determine whether participants acquired influenza due to hospital or community exposure. However, our data on household exposure suggest that such exposures were balanced between intervention groups."} In fact, eq. (2) and eq. (3) are derived with the explicit assumption that transmission in which both groups are wearing the same protection (or lack thereof) will on average be the same. The concern is not solely whether community transmission will be larger in one arm than the other.  Rather, as we have shown quantitatively, even equal community transmission in both groups will bias the point estimate of the risk ratio towards 1, while also giving the illusion that the width of the confidence interval is a factor of $1-b$ smaller than actuality. In other words, the study results are reporting the noise rather than signal. Since there are a larger number of noise events compared to signal events, there is a smaller uncertainty in the average noise, which is the reported observed uncertainty instead of the uncertainty of the signal relevant to the conclusions.

\subsection{The neglect of other sources of data and reasoning}
In contrast to the inadequate evidence from the RCTs conducted thus far, the causal mechanism by which masks prevent infection is understood---if physical exposure to the virus is prevented one will not be infected. Multiple sources of data indicate that masks do in fact filter a substantial fraction of viral particles, especially respirator masks that are tested for this purpose \cite{A1,A2,A3,A4,A5,A6}. The complex question of mask effectiveness, including its dependence on specific conditions of studies, can be framed in terms of the extent to which mask wearing reduces the inhaled and exhaled viral load (i.e. the inhalation and exhalation efficacies) \cite{R7, A4}.\footnote{The theoretical analysis in ref.~\cite{A4} is a special case of our analysis in ref.~\cite{R7}; in particular they ignore threshold effects, which corresponds to our analysis if we constrain $f(v)$ to be a linear function.}  RCTs always require theory to interpret their results (e.g., in our analysis above, we use the theoretical assumption that any effect masks have will occur only while they are worn, see also Appendix D).  When the theoretical assumptions are made more explicit, we see that results of the RCTs and the mechanistic understanding of how masks work are consistent with each other.  

\section{Missing data, compounding and nonlinear effects of masking, including on non-study participants}
Our analysis of the Cochrane review and Loeb 2022 thus far demonstrates that these empirical studies utilize approaches for analysis that make unfounded assumptions in estimating effects. While the methods are conventional, they follow a pro-forma context that is not grounded in the circumstances of the trials and observational studies. In this section we introduce additional errors due to missing data, including on non-study participants, as well as compounding effects.

\subsection{The neglect of source control on transmission}

A mask or respirator that has increased effectiveness for preventing transmission can protect both the mask wearer and others from being infected. Loeb 2022's study of health care workers does not track or report the cases that occur in patients not already infected, nor in the non-participant health care workers, even though they may be infected by participant individuals. The evaluation of an intervention should include the overall relative benefits and harms in scenarios with and without that intervention. The risk ratio for individuals is not well characterized by the Loeb study, as described above, and apparent benefit is further reduced by the absence of an empirical value of a non-measured source control risk, or outcome risk ratio. Indeed, the change in risk for adverse events due to infection---such as death or severe disease of vulnerable patients---may be much higher. Improved source control may reduce the infection rate of multiple other individuals leading to an overall reduction of infections in aggregate. Thus, for example, if the infection reduction is $2$X for one individual, as well as for a second individual due to source control, the reduction in overall number of transmissions is $100\%$ of the number of infections of the individual who changes their mask. Said differently, the same mask efficacy for source control gives $2$X increase in the total number of infections prevented. A similar count for adverse events of vulnerable patients, including death or severe disease, would multiply the number of infections for each type of individual by the probability of a particular adverse event. A direct expression can be written by including all of the infections affected by the individual who has changed mask behavior, showing how the ratio of the entire set of infections is reduced by the wearing of higher quality masks, not just those of the individual who changes masking behavior (see Appendix E). The next section treats the case where risk is affected by changes in masking of both individuals who are in contact.

\subsection{The neglect of the effects of two way masking on transmission}
What happens to risk of infection due to change in mask behavior including source control when essentially everyone changes their masking behavior? For each potential transmission event, there is a reduction of the probability of transmission because of the reduction in viral particle exposures due to source control and due to prevention of inhalation.\footnote{Comparable to studies of efficacy of one-hand clapping, as examples of the general topic of emergent behaviors, where one element depends on the existence of another element for its effect; multicomponent therapies being examples.\cite{f1}} Both of these effects impact all individuals. The impact of both changes of masking is greater than the impact of either alone, and may be multiplicative in the efficacy of masks due to each mask independently reducing the viral concentration by a proportionate amount. Thus, if each of them reduces the risk by a factor of $2$X the overall reduction is a factor of $4$X, and if it is $10$X, as would be consistent with corrected analysis of Loeb 2022, the overall bound on reduction due to wearing N95 masks is not less than $100$X. Absent an RCT empirical bound on source control effects, the available information implies that the bound of uncertainty is not less than $100$X, and perhaps much higher (Appendix F). The estimate would be even larger if hazard ratios for adverse events for vulnerable patients are specified. Ref. \cite{R7} includes a more complete analysis of the nonlinear effects of masking. 

\section[Significance of results for health]{Significance of results for health}

\subsection{Significance and compounding effects}
As previously discussed, the overall impact of an intervention rather than just the risk ratio for an individual must be considered in determining the benefit of an intervention. The example of a bound of a factor of 2X of infections on healthcare workers must first be evaluated in its own right. A 2X bound does not indicate a lack of importance of the intervention unless the reduction does not have a significant effect on the individuals who are involved. A 2X change in transmission can be a significant effect for many contexts of disease prevention and its consequences in adverse events both for healthcare workers \cite{BMA} and for patients \cite{AustraliaHAIdeaths,UKHAI1,UKHAI2}. Moreover, a compounding effect of subsequent transmissions implies that a 2X change in transmission can convert a disease that has a $R_0$ of less than $2$ to an $R$ that is less than one, leading to a reduction of transmission below the endemic threshold for that disease. Any intervention, including vaccination, treatment, or other preventive measure for transmission, that would reduce the transmission rate by 2X would be significant. Our analysis shows that the bound is much higher and therefore can have dramatically greater impact.

\begin{figure}
    \centering
    \includegraphics[width=.935\textwidth]{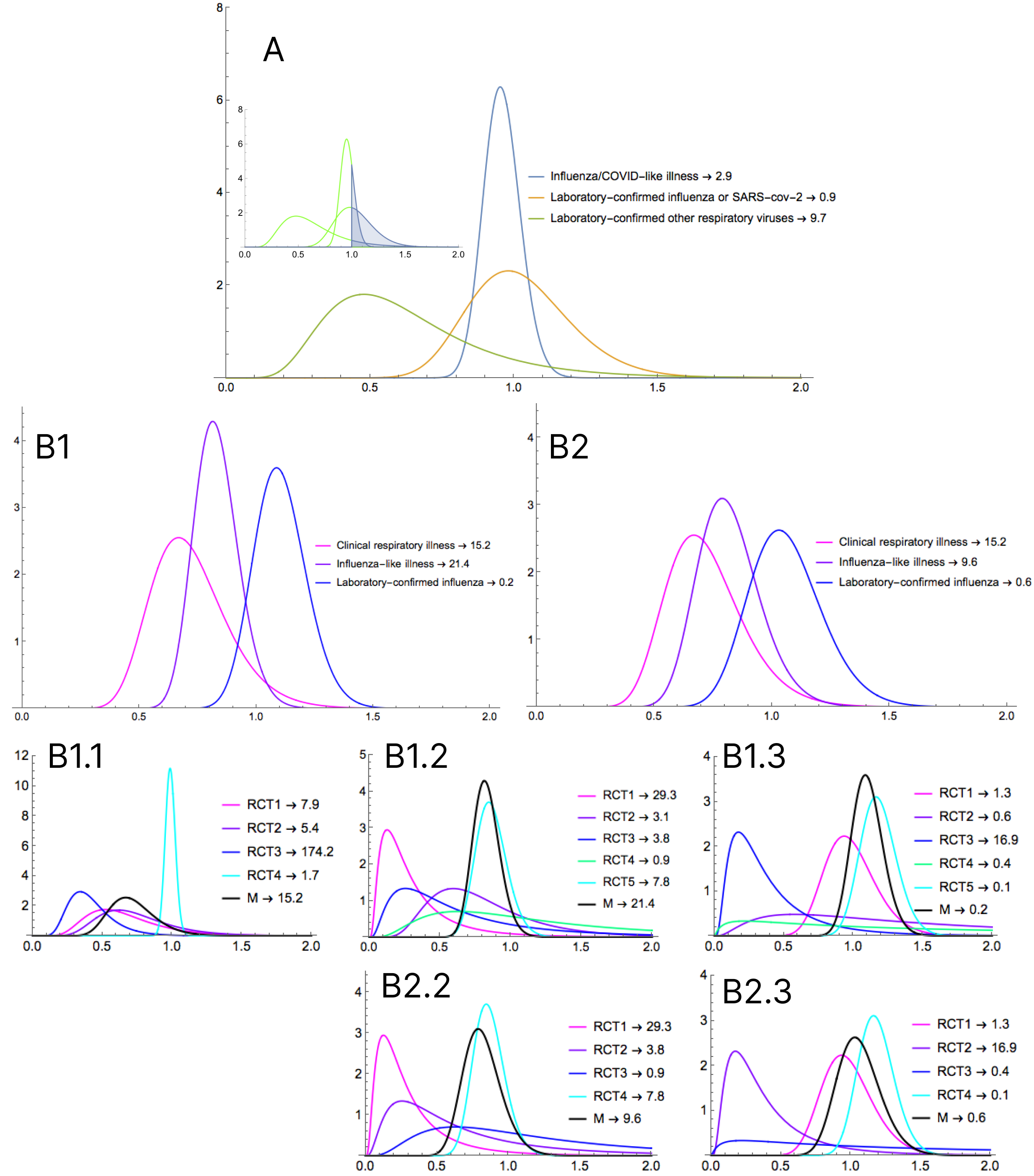}
    \caption{Lognormal distributions obtained from Cochrane review data (see Appendix G) comparing (A) masks and no-masks indicate that the ratio of likelihood for benefit from masks are 2.9 to 1, 0.9 to 1, and 9.7 to 1 (inset shows unshaded as benefit and detriment as shaded)---two out of three indicate benefit for masking by a high probability ratio; (B1) N95 respirators compared with non-respirator masks with benefit of N95s also have 2 out of 3 high benefit likelihood ratios, 15.2, 21.4 and 0.2. The number of studies included in (B1) are 3, 5, and 5, for the end points clinical respiratory illness, influenza-like illness, and laboratory-confirmed influenza, respectively. When one of the 5 of the last two end points is omitted (B2) the ratios are 9.6 and 0.6. Details of the meta analyses (B1.1-B1.3) and (B2.1-B2.3) show study and meta analysis log normal distributions and the relative probability of benefit. One of the studies Radonovich 2019 ref. \cite{R10} shown in light blue dominates the meta analyses but has protocols to put on masks or N95s only within 6 ft of patients, despite such protocols being counter to the principles of airborne transmission. }
    \label{fig:significance}
\end{figure}

\subsection{Significance for health and statistical significance}
The objective of data analysis is to examine the associations between exposure variables of interest and outcomes and to characterize the strength of the associations of these variables with occurrence of the outcomes.  The development of procedures for doing so may be associated with their pro-forma application without regard to the underlying motivation. Indeed, the assessment of statistical significance diverts attention from the analysis of significance construed broadly. We can directly show (see uniform prior Bayesian analysis in Appendix G) that the Cochrane meta-analyses convincingly demonstrate that there is a benefit of N95 respirators over medical masks in clinical settings, and masks over no-masks. We calculate the odds ratio that N95 masks reduce transmission compared to medical masks using the posterior probability distributions for the means obtained from the studies and meta-analyses. For example, \cref{fig:significance}(B1.1) shows that the endpoint of clinical respiratory illness for N95 respirators versus medical masks includes 5 studies and their combination in a meta-analysis. All of the studies have a probability ratio of benefit to detriment that is greater than one, and their combination in the meta-analysis gives a 15 to 1 probability advantage of N95s. Of meta-analyses, two out of three endpoints for masking over no masks, and two out of three endpoints for N95 respirators over masks have large probability of benefit ratios. Thus, the reported meta-analysis results, even with all of the stated limitations of study protocols that reduce effect sizes, nevertheless show that with high probability masking and N95s in particular, as key examples of physical interventions, reduce the spread of respiratory viruses and their adverse effects on health. 

While the benefit found in the studies is clear, we should still consider their statistical significance (this analysis does not depend on Appendix G). One study, Radonovich 2019 \cite{R10}, dominates the comparison of N95s versus masks with over 40\% to nearly 90\% weight in different endpoints. The protocols of this study specify that masks or N95 respirators should be donned within 6 ft of patients. Since airborne transmission occurs within the connected air space to an infected individual (and even after they have departed from that space), this protocol is counter to the principles of N95 respirator use and therefore does not test N95 efficacy.

We performed a meta-analysis omitting Radonovich 2019 and found statistical significance at the $95\%$ level in two out of three meta analyses on N95 versus surgical masks. For clinical respiratory illness we have  a risk ratio and confidence interval $0.53~[0.36,0.74]$, and for influenza-like disease $0.60~[0.35,0.95]$. These can be compared with the Cochrane review results of $0.70~[0.45, 1.10]$ and $0.82~[0.66, 1.03]$. The new results whose upper bound of the 95\% confidence interval is less than one satisfies the usual criterion of statistical significance. The means indicate a factor of approximately two-fold reduction of transmission due to N95 respirators compared to surgical masks, but may be as high as a factor of three.

We note in addition that no sensitivity analyses were reported on masking in the Cochrane review. A sensitivity analysis on a meta analysis by Long 2020 \cite{R8} reported statistical significance for the benefit of N95s versus medical masks when removing one study, Loeb (2009) \cite{R9}, for prevention of laboratory-confirmed respiratory viral infections, and by removing Radonovich 2019 for prevention of respiratory infections. 

\section{Adverse events of masking}
Tracking adverse events is an important part of any clinical study and the implications of adverse events in procedural masks and the N95 respirator group should be understood in terms of their implications for the utilization of both low and high quality masks.\footnote{Benefit and harm should be described in comparable terms so that the overall impact can be characterized or even quantified ($\mathcal{B-H}$).} The Loeb 2022 study summary reports on adverse events as follows: ``There were 47 (10.8\%) adverse events related to the intervention reported in the procedural mask group and 59 (13.6\%) in the N95 respirator group" without further commentary. The text of the study itself does not report on the nature of most of the adverse events, reporting only on 2 participants with serious adverse events in the medical mask group and 1 in the N95 group due to infections, and 2 in the medical mask group and 1 in the N95 group due to need to isolate in a hospital. The remainder, which are mild, are not reported in the paper but in the supplement as otherwise unspecified ``discomfort," ``skin irritation" and ``headaches." As written, the summary might raise alarms, however it is clear from the supplement that the significance of adverse side effects should be clearly stated, not just their numbers. Since the other events did not require hospitalization, the adverse events should have been reported as ``mostly mild." As with any intervention, the prevention of transmission is a major benefit and the absence of severe adverse events except those due to infections that were not prevented informs the utilization of the intervention in any setting.  

\section{Discussion/Summary}

Studies of the efficacy and effectiveness of respiratory protection have potentially profound implications as decisions are made on policies to protect the public generally and workers specifically against airborne infections.  In this paper, we have shown that empirical results, interpreted as showing no or weak effects of proven methods of respiratory protection in the workplace, rest on flawed assumptions in study design and analysis.  Most critically, studies do not reflect the critical distinction between infections resulting from exposure while using an intervention from those resulting from exposures when the intervention is not being used. We show why these empirical results are flawed and the scientifically incorrect conclusions that result.  The net result is a strong bias towards null findings and a failure to reflect the full statistical uncertainty of study results.  These flaws of individual studies are compounded in meta-analyses.   

Conventional analyses do not correct quantitatively for these problems. Technically, the analysis of clinical studies of mask efficacy are missing four things: (1) propagation of uncertainty leading to biased estimate of the mean and underestimate of uncertainty, (2) compounding of effects of masking on non-study participants, (3) importantly, analysis of significance---the meaning of the results through their implications for health (current results, even with all of the limitations that reduce effect sizes, show that with high probability physical interventions reduce the spread of respiratory viruses), and (4) proper contextualization of purported adverse effects of mask wearing. We illustrate these points by reanalyzing both the Cochrane review and a trial that claims to limit the benefits of N95 respirator masks compared to surgical masks in health care workers to 2X, and show that correcting the analysis based upon their own empirical data results in a bound of 10X or even larger. Further, the results only apply to an unspecified class of low risk procedures. Considering the broader benefits of respiratory protection, including the benefit of source control for infections in others, yields an even higher upper bound of potential benefit.  In other words and consistent with the laboratory-established efficacy of N95 respirators, we show that, including the effect of source control on infections of others (including patients), the available empirical evidence is consistent with masks being highly effective.  

The empirical approach taken to date in the design and analysis of studies of respiratory protection has not been adequately grounded in the correct theoretical assumptions around the transmission of airborne infections.  The microenvironmental formulation clarifies the necessity of considering exposures not addressed at all by the intervention(s) of interest.  While we have focused on interventions for respiratory infection, the same considerations can be extended to studies of other interventions, e.g., increased ventilation or ultraviolet radiation.  

For developing guidelines for public health protection, an approach based in gathering all related evidence through systematic review and meta-analysis is generally followed.  However, limitations that extend to an entire body of evidence are not reversed by the review and pooling processes.  We urge caution with using extant interpretations of the evidence on respiratory protection for formulating guidelines. 

\section*{Appendix A}

Given a set of studies of the same quantity with estimates for that quantity $r_i$ and standard error $\sigma(r_i)$ that have biases due to distinct conditions that are unknown, a random effects model assumes that there is an unbiased normally distributed random variable that describes the difference between the studies. An analysis of these assumptions leads to the following approximate equations for combining the studies to achieve an estimate $r$ with standard error $\sigma(r)$ and the related confidence interval $\Gamma_{95\%}$ :
\begin{align}
r = \frac{\sum_i w_i r_i}{\sum_i w_i}
\end{align}
Standard error (use different notation than $\sigma$, so that it can be retained for standard deviation? Or is argument enough)
\begin{align}
\sigma(r) = \sqrt{\frac{\sum_i^{N} \sigma(r_i)^2 (N_i-1)}{\sum_i^N (N_i -1)}}
\end{align}
This can be used for a single study if the study has multiple locations or otherwise has a partition among multiple groups. This does not account for any systematic biases in results as discussed in Appendix A.

Where the value of $w_i$ is given by 
\begin{align}
w_i = \frac{1}{\sigma(r_i)^2 + \tau^2} 
\end{align}
The DerSimonian-Laird value for $\tau^2$ is given by
\begin{align}
\tau^2 = \frac{Q-(N-1)}{e}
\end{align}
where
\begin{align}
Q = \sqrt{\sum_i^{N}\left( \sigma(r_i)^{-2} \left(r_i - \frac{\sum_j^{N} \sigma(r_j)^{-2} r_i}{\sum_j^{N} \sigma(r_j)^{-2}} \right) \right)}
\end{align}
\begin{align}
e =  \sum_i \sigma(r_i)^{-2} -  \frac{\sum_i \left(\sigma(r_i)^{-2} \right)^2}{\sum_i \sigma(r_i)^{-2}}
\end{align}
(there are a set of possible assumptions different from DerSimonian-Laird and no explanation of what is being used in the meta analysis, differences are likely to be small but sometimes they might matter with the level of precision needed to estimate the confidence interval and statistical significance, even numerical rounding error may matter)

For a log normal distribution treatment that is generally used for risk ratios the variables have to be transformed to and from the log normal variables $l_i = \ln{r_i}$ giving the aggregated estimate $l$ and the standard error $\sigma(l)$. (It should not matter that we are using $\ln$ instead of $\log_{10}$, which adds a constant to the transformed variables this should be confirmed).
\begin{align}
l_i = \ln{r_i} - \sigma(r_i)^2/2
\end{align}
\begin{align}
\sigma(l_i) = \sqrt{\frac{\sigma(r_i)^2}{r_i^2} +1}
\end{align}
The back transform, according to a geometric approximation, is given by
\begin{align}
r = e^{l}
\end{align}
and the $95\%$ confidence interval by 
\begin{align}
\Gamma_{95\%}(r) = e^{l\pm 1.96 \sigma(l)}
\end{align}
Other approximations don't appear to give substantially different results but this should also be checked.

\section*{Appendix B}
\begin{figure}
    \centering
    \includegraphics[width=.40\textwidth]{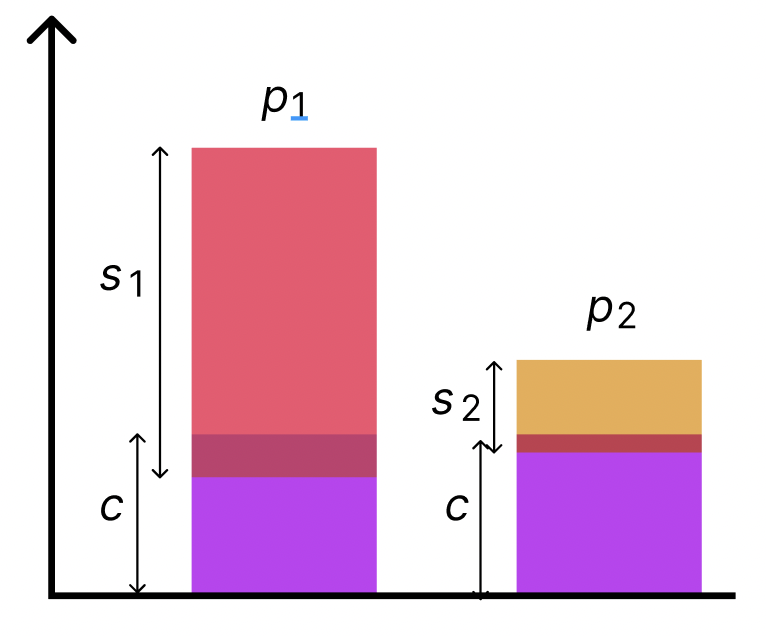}
    \caption{A schematic diagram for possible values of the infection probabilities in the control and intervention groups $p_1$, $p_2$, $c$, $s_1\equiv\frac{1}{N}\sum_{j=1}^N s_j$, and $s_2\equiv \frac{1}{N}\sum_{j=1}^Nrs_j=rs_1$.}
    \label{fig:diagram}
\end{figure}

Let us more carefully consider how we derived \cref{eq:c}.  Suppose that an individual $j$ has a $\hat{c}_j$ chance of infection during the time period for which they are unmasked, and an independent chance $r\hat{s}_j$ of infection during the time period while masked, where $r$ is the actual risk ratio and $\hat{s}_j$ is the chance of infection during that time period were the individual in the control group.  Thus, were this individual in the control group, the chance of infection would be $\hat{c}_j+\hat{s}_j-\hat{s}_j\hat{c}_j$.  In the mask group, this individual's chance of infection is $\hat{c}_j+r\hat{s}_j-r\hat{s}_j\hat{c}_j$.  Assuming that the composition of the control group is on average equivalent to that of the mask group, 
\begin{align}
p_1=&\frac{1}{N}\sum_{j=1}^N \hat{c}_j+\frac{1}{N}\sum_{j=1}^N(\hat{s}_j-\hat{s}_j\hat{c}_j) \\
p_2=&\frac{1}{N}\sum_{j=1}^N \hat{c}_j+r\frac{1}{N}\sum_{j=1}^N(\hat{s}_j-\hat{s}_j\hat{c}_j)
\end{align}
where $N$ is the number of individuals in the mask group (these equations are illustrated in \cref{fig:diagram}).  Letting 
\begin{equation}
c\equiv\frac{1}{N}\sum_{j=1}^N \hat{c}_j
\end{equation}
 be the average probability of unmasked infection in the mask group, we see that 
 \begin{equation}
 \frac{1}{N}\sum_{j=1}^N (\hat{s}_j-\hat{c}_j\hat{s}_j)=p_1-c
 \label{eq:p1c}
 \end{equation}
 and thus $p_2=c+r(p_1-c)$, reproducing \cref{eq:c}.  
 
 We note that $b=c/p_1$ will be larger than the fraction of exposures for which individuals are unmasked:
 \begin{equation}
 \frac{c}{p_1}>\frac{c}{p_1+\frac{1}{N}\sum_{j=1}^N \hat{c}_j\hat{s}_j}=\frac{c}{c+\frac{1}{N}\sum_{j=1}^N \hat{s}_j}
 \end{equation} 
(The total exposure is $c+\frac{1}{N}\sum_{j=1}^N \hat{s}_j$, and thus $\frac{c}{c+\frac{1}{N}\sum_{j=1}^N \hat{s}_j}$ is the fraction of exposures for which individuals in the intervention group are unmasked.  The equality follows from \cref{eq:p1c}.)
 
 \section*{Appendix C}
In Loeb 2022 \cite{R9}, they used survival analysis, which allows for individuals to be  removed from the population after infection and also allows for the infection rates to vary with time.  However, the hazard ratio $p_1(t)/p_2(t)$ of the infection rates of the two groups is assumed to be time-independent. 

 Let $p_1(t)$ be expected rate of infection in the surgical mask group, $p_2(t)$ be the expected rate of infection in the N95 respirator group, and $c(t)$ be the expected rate of infection from community transmission, which is assumed to be the same for both groups.  Then we can write
\begin{equation}
p_2(t)=\left(p_1(t)-c(t)\right)r+c(t)
\end{equation}
where $r$, which we assume to be time-independent, is the actual hazard ratio of N95 respirators to surgical masks.  Assuming that $b=c(t)/p_1(t)$ is independent of time yields 
\begin{equation}
r=\frac{p_2(t)/p_1(t)-b}{1-b}
\label{eq:nb}
\end{equation}
The confidence intervals in the study are given for the
hazard ratio $p_1(t)/p_2(t)$ (assumed to be time-independent), and so can be transformed according to \cref{eq:nb} (see right panel of \cref{fig:surgicalandN95}).  Note that for the sake of consistency with our analysis of surgical masks vs. no masks, the hazard ratio $r$ considered here is for N95 respirators to surgical masks, which is the inverse of the hazard ratio for surgical masks to N95 respirators that is considered in the original study.

 \section*{Appendix D}

Due to the ongoing lack of clarity about the science of mask wearing (and other precautions) here is a summary of essential points that are consistent with the science of mask wearing and its prevention of transmission. The essential property of wearing a mask or other precautions is that the effectiveness is measured by preventing a single infection and not by the rate of infections (generalization to the rate of infections is possible). This applies in modified form to the consideration of masks versus no masks, and respirator versus surgical masks. The essential question is whether prevention of transmission occurs due to the adoption of the particular measure. In every case, ``wearing a mask" should be understood to be following protocols consistent with its function, such as covering mouth and nose, adequate fit for respirator masks, wearing within the shared physical space and time interval in which viral particles propagate not just within a limited distance of a potentially infected individual for airborne transmission. 

\section*{Appendix E}
 
An individual's mask behavioral choice does not only affect their own infections but also all of the other individuals who are infected by them. To formalize the infection rate change for both the individual who changes their mask and for others due to source control, we can evaluate the overall reduction in relation to the total number of infections due to the behavioral choice of a single individual. The reduction of the infection can be written as

\begin{equation}
r = \frac{u \sum_j T_{i,j} + v \sum_j T_{j,i}}{\sum_j T_{i,j} + \sum_j T_{j,i}}
\end{equation}
where $T_{i,j}$ is the reference probability of individual $i$ being infected by individual $j$, $u$ is the reduction of transmission to the individual wearing the improved mask, and $v$ is the reduction of transmission to others (source control). 
This expression manifests the rates of infection of the individual who changes mask behavior, with the average effect characterized by $u$, as well as the individuals who are around them, with the average effect characterized by $v$. The effects may be different for multiple reasons, including the degree of asymptomatic transmission for different diseases, mechanism of transmission (e.g. droplet or aerosol) or other behavioral choices such as quarantine due to exposure, or isolation during symptomatic infection. For the case where both $u$ and $v$ are similar, a change of effectiveness of a mask by an individual by 2X or by 10X results in a proportional change in the infections that that person experiences due to other individuals, as well as those that are due to improved source control. The same risk ratio reduction has a much larger impact on the number of infections that occur than considering only one or the other. A more general expression with similar conclusions can be written for adverse events such as deaths or severe disease that are higher for higher vulnerability individuals, including patients, as
\begin{equation}
r = \frac{u w_i \sum_{j} T_{i,j} +  v \sum_{j} w_j T_{j,i}}{w_i \sum_{j} T_{i,j} + \sum_{j} w_j T_{j,i}}
\end{equation}
where distinct outcome vulnerabilities $w_i$ and $w_j$ of the individual who is changing their mask behavior and of others are specified. In this context, the infection of others by healthcare workers also evokes professional responsibility, patient rights, and informed consent.

\section*{Appendix F}

Two-way masking, the effect of individuals masking who are both infectious and infected, results in higher levels of protection than one or the other masking alone. More generally, specific individuals masking is less effective protection than universal masking. 

For a single exposure event (which we define as exposure over a small enough time window that we can ignore time effects), suppose that the probability of infection is $p(v)$ where $v$ is the viral load.  Following the analysis of ref. \cite{R7} we can define $f(v)\equiv -\ln(1-p(v))$ such that $p(v)=1-e^{-f(v)}$.    Now suppose we wish to consider a bunch of exposure events with viral loads $v_1, v_2,....$ that are each separated enough in time that the probability of being infected by any of them is independent.   Then the overall probability of infection can be written as 
\begin{equation}
p=1-\Pi_n (1-p(v_n))=1-\Pi_n e^{-f(v_n)}=1-e^{-\tilde v}
\end{equation}
where $\tilde v = \sum_n f(v_n)$ is the total effective exposure.    For any particular event $n$, the associated effective exposure is $\tilde v_n=f(v_n)$ but of course we can treat $\tilde v_n$ as an effective parameter that could depend on other factors besides just $v_n$ (i.e. we can simply define $\tilde v_n\equiv -\ln (1-p_n)$ where $p_n$ is the probability of being infected for that event.) 

In any case, if we do wish to consider the function $f(v)$, we note that a mask that reduces $v$ by a factor of $2$ could reduce the $f(v)$ by a greater factor.   For instance, if $f(v)=v^\beta$, then the mask will reduce effective exposure by $2^\beta$. In the absence of threshold effects, $\beta =1$, with larger values of $\beta$ corresponding to increasingly strong threshold effects which affect the infectious rate when the viral exposure is small. For general forms of $f(v)$ and in the more general case where the effective exposure depends on more than just the viral load, the degree to which a mask reduces effective exposure will vary but one should be able to choose some effective parameter for mask efficacy that will work.  

We now show that for any particular value of the effective exposure, the risk ratio of everyone masking is less than the square of only the susceptible individuals masking. For any particular value of the effective exposure $\tilde v$,  if we assume that one person masking reduces effective exposure $\tilde v \rightarrow b \tilde v$ and both people masking reduces effective exposure $\tilde v \rightarrow b^2 \tilde v$, then we have that the risk ratio for one-mask is 
\begin{equation}
\frac{p'}{p}=\frac{1-e^{-b\tilde v}}{1-e^{-\tilde v}}
\end{equation}
while if everyone is wearing a mask, it's 
\begin{equation}
\frac{p''}{p}=\frac{1-e^{-b^2\tilde v}}{1-e^{-\tilde v}}
\end{equation}

Consider the function $h(x)=\ln(1-e^{-e^x})$.   One can verify that this function is concave.   Thus, 
\begin{equation}
h(\ln v+2\ln b)-h(\ln v) \leq 2(h(\ln v + \ln b) -h(\ln v)).
\end{equation}
This implies that for any particular value of the effective exposure
\begin{equation}
\frac{p''}{p} \leq \left(\frac{p'}{p}\right)^2.
\end{equation}
which is the desired result that the risk ratio for two-way masking (on the left) is lower than the square of the risk ratio for one-way masking (on the right).

Thus, the reduction in risk for the reduced viral exposure due to universal masking can have a multiplicative effect, and even more due to both convexity of the probability of infection (arising from a shift from being near to the high exposure limit, to a lower exposure in the control regime), and to threshold effects that may arise at low levels of exposure that make exposures even less likely to infect.

\section*{Appendix G}

In this Appendix we re-analyze the data from a Bayesian perspective.   Consider a control group and intervention group with $N_1$ and $N_2$ individuals and assume that each individual in group $i$ has an independent probability of infection $p_i\in[0,1]$.  We assume a maximum-entropy prior: 

\begin{equation}
\pi(p_1,p_2)=\begin{cases} 1 \text{~~if~~} p_1,p_2\in[0,1] \\ 0 \text{~~otherwise} \end{cases}
\end{equation}
In other words, the prior for each $p_i$ is independent of the other and uniform over $[0,1]$.   

Using the fact that a uniform distribution over $[0,1]$ is equivalent to a beta distribution $\text{Beta}(\alpha=1,\beta=1)$ and that a beta prior is conjugate to the Bernoulli likelihood function, the posterior distribution over each $p_i$ is independent of the other and equal to $\text{Beta}(\alpha=1+n_i,\beta=1+N_i-n_i)$ where $n_i$ is equal to the number of observed infections in group $i$.   

For large $\alpha$ and $\beta$, i.e. for large values of $n_i$ and $N_i-n_i$, the beta distribution can be approximated by a normal distribution with the same mean and variance.   In this limit, we replace $\alpha=1+n_i$ with $\alpha=n_i$ and $\beta=1+N_i-n_i$ with $\beta=N_i-n_i$, which gives a mean and variance of $n_i/N_i$ and $n_i(N_i-n_i)/N_i^3$, respectively.   We note that the mean and variance are equivalent to the empirical mean $\hat p_i=n_i/N_i$ and the empirical variance $\hat \sigma_i^2=\hat p_i(1-\hat p_i)/N_i$.   

Thus under the assumption of a uniform prior, the posterior for each $p_i$ is independently given by $\pi(p_i|\hat p_i)\sim N(\hat p_i,\hat \sigma_i^2)$.   In order to obtain an approximate distribution for the posterior of the risk ratio $r=p_2/p_1$, we approximate the normal distributions of $p_1$ and $p_2$ with lognormal distributions and then use the fact that the ratio of two log-normally distributed random variables is also log-normal.

 \end{document}